\documentclass[aps,prd,twocolumn,superscriptaddress]{revtex4-1}%
\usepackage[utf8]{inputenc}
\usepackage{dsfont}
\usepackage{IEEEtrantools}
\usepackage{amsfonts}
\usepackage{amsmath}
\allowdisplaybreaks[4]        
\usepackage{amssymb}
\usepackage{euscript}     
\usepackage{lipsum}[1]
\usepackage{braket}
\usepackage{starfont}
\usepackage{color,soul}         
\usepackage{tensor}        
\usepackage{amsthm}
\usepackage{graphicx}
\usepackage{slashed}
\usepackage{leftidx}
\usepackage{subfigure}
\usepackage{bbm}
\definecolor{outerspace}{rgb}{0.25, 0.29, 0.3}
\definecolor{scarlet}{rgb}{1.0, 0.13, 0.0}
\usepackage[header,title,page,titletoc]{appendix}  
\definecolor{princetonorange}{rgb}{1.0, 0.56, 0.0}
\definecolor{WildStrawberry}{rgb}{1.0, 0.26, 0.64}
\definecolor{rossocorsa}{rgb}{0.83, 0.0, 0.0}
\definecolor{navyblue}{rgb}{0.0, 0.0, 0.5}
\usepackage{float}
\usepackage[paper=letterpaper,margin=1in]{geometry}
\usepackage{mathtools}

\newcommand{\labell}[1]{\label{#1}} 

\DeclareMathAlphabet{\pazocal}{OMS}{zplm}{m}{n}
\newcommand{\req}[1]{(\ref{#1})} 

\newcommand{\bea}{\begin{eqnarray}}
\newcommand{\diff}{\mathrm{d}}
\newcommand{\eea}{\end{eqnarray}}
\newcommand{\ba}{\begin{eqnarray}}
\newcommand{\ea}{\end{eqnarray}}

\newcommand{\be}{\begin{equation}}
\newcommand{\ee}{\end{equation} }
\newcommand{\beqa}{\begin{eqnarray}}
\newcommand{\eeqa}{\end{eqnarray}}
\newcommand{\beqar}{\begin{eqnarray*}}
\newcommand{\eeqar}{\end{eqnarray*}}

\renewcommand{\req}[1]{eq.~(\ref{#1})}

\newcommand{\ssc}{\scriptscriptstyle}
\newcommand{\eg}{{\it e.g.,}\ }

\renewcommand{\c}{$c$}

\makeatletter
\newcommand{\dal}{\mathop{\mathpalette\dal@\relax}}
\newcommand{\dal@}[2]{%
  \begingroup
  \sbox\z@{$\m@th#1\square$}%
  \dimen0=\fontdimen8
    \ifx#1\displaystyle\textfont\else
    \ifx#1\textstyle\textfont\else
    \ifx#1\scriptstyle\scriptfont\else
    \scriptscriptfont\fi\fi\fi3
  \makebox[\wd\z@]{%
    \hbox to \ht\z@{%
      \vrule width \dimen0
      \kern-\dimen0
      \vbox to \ht\z@{
        \hrule height \dimen0 width \ht\z@
        \vss
        \hrule height 2\dimen0
      }%
      \kern-2.5\dimen0
      \vrule width 2.5\dimen0
    }%
  }%
  \endgroup
}
\makeatother

\usepackage{hyperref}
\hypersetup{
    colorlinks,
    citecolor=rossocorsa,
    filecolor=navyblue,
    linkcolor=navyblue,
    urlcolor=navyblue
}
\begin{document}

\title{Regular black holes in three dimensions}
\author{Pablo Bueno}
\affiliation{Instituto Balseiro, Centro At\'omico Bariloche. 8400-S.C. de Bariloche, R\'io Negro, Argentina.}

\author{Pablo A. Cano}
\affiliation{Instituut voor Theoretische Fysica, KU Leuven. Celestijnenlaan 200D, B-3001 Leuven, Belgium. }

\author{Javier Moreno} 
\affiliation{Instituto de F\'isica, Pontificia Universidad Cat\'olica de Valpara\'iso. Casilla 4059, Valpara\'iso, Chile.  }
\affiliation{Center for Quantum Mathematics and Physics (QMAP). Department of Physics \& Astronomy, University of California, Davis, CA 95616 USA.}

\author{Guido van der Velde}
\affiliation{Instituto Balseiro, Centro At\'omico Bariloche. 8400-S.C. de Bariloche, R\'io Negro, Argentina.}


\begin{abstract}

We find a plethora of new analytic black holes and globally regular horizonless spacetimes in three dimensions. The solutions involve a single real scalar field $\phi$ which always admits a magnetic-like expression proportional to the angular coordinate. The new metrics, which satisfy $g_{tt}g_{rr}=-1$ and represent continuous generalizations of the BTZ one, solve the equations of Einstein gravity corrected by a new family of densities (controlled by unconstrained couplings) constructed from positive powers of $(\partial \phi)^2$ and certain linear combinations of $R^{ab} \partial_a \phi \partial_b \phi$  and $(\partial \phi)^2 R$. Some of the solutions obtained describe black holes with one or several horizons. A set of them possesses curvature singularities, while others have conical or BTZ-like ones. Interestingly, in some cases the black holes have no singularity at all, being completely regular. Some of the latter are achieved without any kind of fine tuning or constraint between the action parameters and/or the physical charges of the solution. An additional class of solutions describes globally regular and horizonless geometries. 


\end{abstract}
\maketitle


\section{Introduction}
Since the seminal discovery of the BTZ solution \cite{Banados:1992wn,Banados:1992gq}, the catalog of three-dimensional black holes has grown in different directions. On the one hand, higher-curvature modifications of Einstein's theory such as New Massive Gravity \cite{Bergshoeff:2009hq} and its extensions \cite{Gullu:2010pc,Sinha:2010ai,Paulos:2010ke} allow for new solutions which differ from the BTZ one \eg in being locally inequivalent from AdS$_3$, in possessing asymptotically flat, dS$_3$ or Lifshitz  asymptotes, or in including curvature (rather than conical or BTZ-like) singularities \cite{Bergshoeff:2009aq,Oliva:2009ip,Clement:2009gq,Alkac:2016xlr,Barnich:2015dvt,AyonBeato:2009nh,Gabadadze:2012xv,Ayon-Beato:2014wla,Fareghbal:2014kfa,Nam:2010dd,Gurses:2019wpb,Gabadadze:2012xv}. Including extra fields also allows for progress, and additional solutions are known for Einstein-Maxwell \cite{Clement:1993kc,Kamata:1995zu,Martinez:1999qi,Clement:1993kc,Hirschmann:1995he,Cataldo:1996yr,Dias:2002ps,Cataldo:2004uw,Cataldo:2002fh} as well as for Einstein-Maxwell-dilaton \cite{Chan:1994qa,Fernando:1999bh,Chen:1998sa,Koikawa:1997am} and Maxwell-Brans-Dicke type \cite{Sa:1995vs,Dias:2001xt}  theories. These typically include logarithmic profiles for some of the fields and curvature singularities. Black hole solutions for minimally and non-minimally coupled scalars have also been constructed \cite{Martinez:1996gn,Henneaux:2002wm,Correa:2011dt,Zhao:2013isa,Tang:2019jkn,Karakasis:2021lnq}, including some which exploit well-defined limits of Lovelock theories to three dimensions \cite{Hennigar:2020drx,Hennigar:2020fkv,Ma:2020ufk,Konoplya:2020ibi}. These typically contain curvature singularities and sometimes globally regular scalars.
A final class involves coupling Einstein gravity to non-linear electrodynamics \cite{Cataldo:1999wr,Myung:2008kd,Mazharimousavi:2011nd,Mazharimousavi:2014vza,Hendi:2017mgb}. For these, examples of regular black holes have been presented for special choices of the modified Maxwell Lagrangian  \cite{Cataldo:2000ns,He:2017ujy,HabibMazharimousavi:2011gh}. 

In this letter we extend the above catalog with a new family of three-dimensional modifications of Einstein gravity involving a non-minimally coupled scalar which admit a large family of analytic generalizations of the BTZ metric describing black holes and globally regular horizonless spacetimes. The new black holes display one or several horizons and include curvature, conical, BTZ-like or no singularity at all depending on the case.


The latter class is particularly interesting. Indeed, the resolution of black hole singularities is an expected property of UV-complete theories of gravity. However, despite numerous efforts at characterizing regular black holes \cite{bardeen1968non,Dymnikova:1992ux,Borde:1996df,Hayward:2005gi,AyonBeato:1998ub,AyonBeato:1999ec,AyonBeato:1999rg,Balakin:2007am,Frolov:2014jva,Frolov:2016pav,Lemos:2011dq,Biswas:2011ar,Olmo:2012nx,Balakin:2015gpq,Olmo:2015axa,Fan:2016hvf,Chamseddine:2016ktu,Menchon:2017qed,Carballo-Rubio:2018pmi,Simpson:2018tsi,Simpson:2019mud,Sert:2015ykz,Buoninfante:2018xiw,Buoninfante:2018stt,Giacchini:2018gxp,Cano:2018aod,Hayward:2005gi,Estrada:2020tbz,Baake:2021jzv,Babichev:2020qpr}, the success in the construction of actual solutions for explicit models which do not rely on the introduction of {\it ad hoc} matter nor require an important degree of fine tuning in the Lagrangian has been moderate at most. 

Our regular black holes come in two different groups. For both, the action takes a simple polynomial form, the couplings are completely unconstrained and so is the mass of the solutions. Naturally, as corresponds to regular black holes, the curvature invariants are finite everywhere. Now, for the first type of solutions, corresponding to metric factors $f(r)\equiv -g_{tt}=g_{rr}^{-1}$ which satisfy $f(r=0)=1$, there is a single constraint which fixes the  ``magnetic charge'' associated to the scalar field in terms of the action couplings ---see \req{sing1} and \req{sing2} below for an explicit example. For the second type, on the other  hand, the metric functions behave as  $f\overset{r \rightarrow 0}{\longrightarrow} \mathcal{O}(r^{2s})$ with $s\geq 1$, and the solutions arise without imposing any constraint whatsoever between the action parameters and/or physical charges of the solution ---see \eg \req{sing3} and \req{frl} for an explicit case.

Our construction here relies on a set of ideas that have been successfully applied to the construction of black hole solutions in higher-dimensions \cite{Quasi,Quasi2,Dehghani:2011vu,PabloPablo,Hennigar:2016gkm,PabloPablo2,Hennigar:2017ego,Hennigar:2017umz,PabloPablo3,Ahmed:2017jod,Feng:2017tev,PabloPablo4,Bueno:2019ycr,Cisterna:2017umf,Bueno:2019ltp,Cano:2020ezi,Cano:2020qhy,Frassino:2020zuv,KordZangeneh:2020qeg}. The theories involved possess particularly simple static black hole solutions, characterized by a single function which, in some cases, satisfies an algebraic equation. While the approach was first developed for purely gravitational theories, certain additional fields turn out to fit naturally within the framework. In four dimensions, the natural choice is a ``magnetic ansatz'' for a Maxwell field, which automatically satisfies the equations of motion of general non-minimally coupled higher-curvature theories and at the same time has a non-trivial effect on the metric  ---yet simple enough to allow for fully analytic solutions \cite{Cano:2020ezi,Cano:2020qhy}. In three dimensions, the analogous magnetic r\^ole is played by $\partial_a \phi$ ---where $\phi$ is a real scalar--- which is related to the Maxwell field strength via a duality transformation. Hence, we construct our actions considering Lagrangians of the form $\mathcal{L}[g^{ab},R_{ab},\partial_a\phi]$, asking them to allow for single-function metrics, $g_{tt} g_{rr}=-1$, and assuming a simple magnetic ansatz for $\phi$, proportional to the angular coordinate. This selects a rather broad class of theories which we denote three-dimensional ``Electromagnetic Quasitopological'' (EM-QT) gravities, following the analogy with higher dimensions\footnote{These are not to be confused with the so-called ``Quasitopological electromagnetism'' theories \cite{Liu:2019rib,Cisterna:2020rkc}.}.

The structure of the letter is as follows. In Section \ref{EMQT} we present our new class of modifications of Einstein gravity as well as the general result for the metric function for a static and spherically symmetric ansatz. In Section \ref{dualf} we comment on the electromagnetic dual frame of our new theories, which involves a Maxwell field rather than a scalar and present the general solution for the  corresponding electric potential. In Section \ref{blackholes} we analyze the different types of spacetimes described by the general solution, emphasizing their horizon and singularity structure. In Section \ref{rotat} we show how to generalize our solutions to rotating ones. We conclude in Section \ref{finalco} with some final comments and future directions. In Appendix \ref{eqsofmo} we present the full non-linear equations of motion of the new theory as well as an equivalent on-shell approach which allows for a simpler extraction of the metric function equation.

\vspace{-0.2cm}
\section{Electromagnetic quasitopological gravities in three dimensions}\label{EMQT}
Inspired by their higher-dimensional counterparts \cite{Cano:2020ezi,Cano:2020qhy}, we define ``Electromagnetic Generalized Quasitopological gravities'' (EM-GQT)  in three dimensions by the condition that a general Lagrangian $\sqrt{|g|}\mathcal{L}[g^{ab},R_{ab},\partial_a \phi]$  becomes a total derivative when evaluated on the ansatz\footnote{For discussions on the general properties of static and spherically symmetric metrics satisfying the $g_{tt} g_{rr}=-1$ condition, see \eg \cite{Podolsky:2006du,Jacobson:2007tj,Hervik:2019gly}.}
\begin{equation}\label{eq:singlefmetric}
\diff s^2=-f(r)\diff t^2+\frac{\diff r^2}{f(r)}+r^2\diff \varphi^2\, ,\quad \phi=p \varphi\, ,
\end{equation}
where $p$ is an arbitrary dimensionless constant.
Theories satisfying such a property allow for solutions of the form \req{eq:singlefmetric}, where the equation of $f(r)$ can be integrated once yielding a differential equation of order 2 (at most). In some cases, the integrated equation satisfied by $f(r)$ is in fact algebraic. In that case, which is the one of interest in the present letter, the theories are called ``Electromagnetic Quasitopological'' (EM-QT). We find that the following family of densities belongs to the three-dimensional EM-QT class\footnote{As a matter of fact, one could consider an even more general density of the form 
\begin{align}
 \tilde{\mathcal{Q}}  = K(X)+X F(X) R  -[3F(X)+2X F'(X)]R^{ab}\partial_a\phi\partial_b\phi \, ,
\end{align}
where $X\equiv (\partial\phi)^2$ and $K$ and $F$ are two arbitrary differentiable functions.  }
\begin{equation}\label{EQTG}
I_{\rm \ssc EMQT}=\frac{1}{16\pi G}\int \diff ^3x\sqrt{|g|}\left[R+\frac{2}{L^2}- \mathcal{Q} \right]\, ,
\end{equation}
where
\begin{align} \notag
 \mathcal{Q}  \equiv & \sum_{n=1} \alpha_n L^{2(n-1)} (\partial \phi)^{2n}-\sum_{m=0} \beta_m L^{2(m+1)}(\partial \phi )^{2m}  \\ &\cdot   \left[ (3+2m) R^{bc} \partial_b \phi \partial_c \phi- (\partial \phi )^2 R \right] \, ,
  \end{align} 
and  where we used the notation $(\partial \phi)^2 \equiv (g^{ab}\partial_a \phi \partial_b \phi)$.
In this expression, the $\alpha_n$, $\beta_m$ are arbitrary dimensionless constants. Observe that $\mathcal{Q}$ contains terms which are at most linear in Ricci curvatures. We have found no evidence for the existence of additional EM-QT densities when higher powers of $R_{ab}$ are included.


As anticipated, and explained in more detail in appendix \ref{eqsofmo}, the full non-linear equations of (\ref{EQTG}) evaluated for an ansatz of the form (\ref{eq:singlefmetric}) turn out to reduce to a single independent equation for the metric function $f(r)$, which can be integrated once and solved. The result reads
\begin{widetext} 
\begin{equation}\label{fgen}
f(r)= \left[\displaystyle \frac{r^2}{L^2}-\mu-\alpha_1 p^2 \log (r/L)+ \sum_{n=2} \frac{\alpha_n  p^{2n} L^{2(n-1)}}{2(n-1)r^{2(n-1)}} \right] \cdot \left[ \displaystyle 1+ \sum_{m=0} \frac{ \beta_m p^{2(m+1)} (2m+1) L^{2(m+1)}}{r^{2(m+1)}} \right]^{-1}\, .
\end{equation}
\end{widetext}

Here, $\mu$ is an integration constant related to the mass of the solutions, which is given by
$
M= \mu +\beta_0 p^2+ \alpha_1 p^2 \log(r_0/L)\, ,
$
where $r_0$ is a cutoff radius. Indeed, the total mass is divergent for $r_0\rightarrow\infty$ whenever $\alpha_1\neq 0$, just like in the charged BTZ solution \cite{Martinez:1999qi}. 
We emphasize that \req{fgen} is the only static and spherically symmetric solution of \req{EQTG}. If one considers a more general ansatz, the equations of motion automatically impose the $g_{tt}g_{rr}=-1$ condition ---see appendix \ref{eqsofmo}.

Naturally, if we set all the $\alpha_n$ and the $\beta_m$ to zero, we are left with the usual static BTZ metric with mass $\mu$. Similarly, when only $\alpha_1$ is active, the metric takes the same form as for the charged BTZ black hole \cite{Clement:1993kc,Martinez:1999qi}, a fact which follows from the electromagnetic-dual description of our EM-QT action, on which we comment next. Since the couplings $\alpha_n$ and $\beta_m$ can be tuned at will, our solutions represent continuous multiparametric generalizations of the BTZ metric. 

\section{Dual frame}\label{dualf}
General three-dimensional theories of the form $\mathcal{L}[g^{ab},R_{ab},\partial_a\phi]$ are dual to theories with an electromagnetic field, $\mathcal{L}_{\rm dual}[g^{ab},R_{ab},F_{ab}]$, where $F_{ab}\equiv 2\partial_{[a}A_{b]}$. In particular, the dual field strength is defined by
\begin{equation}
F_{ab}=-\frac{1}{2}\epsilon_{abc}\frac{\partial \mathcal{L}}{\partial\left(\partial_{c}\phi\right)},
\end{equation}
and note that, in this way, the Bianchi identity of $F$ is equivalent to the equation of motion of $\phi$.

 In order to obtain the dual Lagrangian, $\mathcal{L}_{\text{dual}}\equiv \mathcal{L}-F_{ab}\partial_{c}\phi\epsilon^{abc}$, one needs to invert the above relation to get $\partial\phi(F)$. Unfortunately, this cannot be done explicitly for the theories we are considering here. Nevertheless, we can perform a perturbative expansion in powers of $L$, in whose case the dual action reads (for $\alpha_1\neq0$)
\begin{align}
&\mathcal{L}_{\text{dual}} =R+\frac{2}{L^2}-\frac{1}{2\alpha_1} F^2  \\ \notag
& +L^2\left[-\frac{\alpha_2}{4\alpha_1^4}(F^2)^2+3\frac{\beta_0}{\alpha_1^2}\tensor{F}{_{a}^{c}}F^{ab}R_{\langle cb\rangle}\right]+\mathcal{O}(L^4)\, ,
\end{align}
where we denoted $F^2\equiv F_{ab}F^{ab}$ and
$R_{\langle cb\rangle}$ is the traceless part of the Ricci tensor. We stress that the dual Lagrangian has an infinite number of terms even when the original action only has a finite number of them (except when the only non-vanishing coupling is $\alpha_1$).

Observe that the leading term is the usual Maxwell piece, which explains the match with the charged BTZ metric when only  $\alpha_1$ is active. Indeed, solutions of $\mathcal{L}_{\rm \ssc EMQT}$ are also solutions of $\mathcal{L}_{\text{dual}}$. In the dual frame, the original ``magnetically charged'' solutions become ``electrically charged'', with a field strength $F=-(\partial A_t(r)/\partial r) \, \diff t\wedge \diff r$, where $A_t(r)$ is the electrostatic potential. Remarkably, this quantity can be obtained exactly and it reads
\begin{align}\notag
&A_t(r)=-\alpha_1 p \log (r/L)+\sum_{n=2}\frac{n \alpha_n p}{2(n-1)}\left(\frac{L p}{r}\right)^{2(n-1)} \\ \label{at} &+f'(r) L\sum_{m=0}\beta_m (m+1)\left(\frac{L p}{r}\right)^{2m+1}\, ,
\end{align}
where $f(r)$ is given by \req{fgen}. Hence, we can think of our new solutions as ``magnetic'' or ``electric'' depending on which frame we consider. For the former, the action is simple, taking the form (\ref{EQTG}) and the extra field is a scalar whose equation is solved by $\phi=p \varphi$. For the latter, the action can only be accessed perturbatively and the auxiliary field is a standard $1$-form whose equation is solved by $A_t(r)$ as given in  \req{at}. Given the trivial behavior of $\phi$ in the magnetic frame, from now on we will exclusively analyze the metric behavior ---which is the same in both frames--- leaving a more exhaustive analysis of the behavior of  $A_t(r)$ and the electric frame for future work.




\section{Black holes}\label{blackholes}
Depending on the values of the $\alpha_n$ and the $\beta_m$, \req{fgen} describes different kinds of solutions, which we study in the following subsections. Let us start with some general comments.
Firstly, the number of horizons  depends on the number of positive roots of the equation 
\begin{equation}
\frac{r^2}{L^2}-\mu-\alpha_1 p^2 \log \left(\frac{r}{L}\right)+ \sum_{n=2} \frac{\alpha_n p^{2n}L^{2(n-1)}}{2(n-1)r^{2(n-1)}}=0\, ,
\end{equation}
which in turn depends on the values and signs of the $\alpha_n$. 
Turning off all the $\alpha_n$, the solution describes a black hole with a single horizon of radius $r=L \sqrt{\mu}$ whenever $\mu >0$, analogously to the neutral BTZ case. 

 In the next-to-simplest case, corresponding to $\alpha_2 \neq 0$ and $\alpha_1=\alpha_{n\geq 3}=0$, at least one horizon exists as long as $2 p^4 \alpha_2 /\mu^2 \leq 1$ and $\mu >0$. In addition, if $ \alpha_2> 0$, the solution possesses two horizons, except for the case $\alpha_2=   \mu^2/(2p^4)$, which corresponds to an extremal black hole. When they exist, the outer and inner horizons correspond, respectively, to
\begin{equation}\label{rpm}
r_{\pm} = \frac{L \sqrt{\mu}}{\sqrt{2}} \sqrt{1 \pm \sqrt{1- \frac{2 p^4 \alpha_2 }{\mu^2} } }\, .
\end{equation}

If we turn on $\alpha_1$ and turn off the rest of $\alpha_n$, the numerator in the metric function is identical to the whole $f(r)$ corresponding to the charged BTZ metric with the identification $Q^2\equiv 2\alpha_1 p^2$. In that case, the horizons structure was studied in detail in \cite{Martinez:1999qi}. Whenever $\mu>1$ the metric describes a black hole with two horizons. For $0<\mu<1$, we have black holes for $2\alpha_1 p^2 \leq Q_1^2$ and for $2\alpha_1 p^2 \geq Q_2^2$ where $Q_{1,2}$ are the roots of the equation $\mu=(Q/2)^2[1-\log (Q /2)^2]$. Whenever, the inequalities are saturated, the black holes are extremal. Interestingly, even for negative values of $\mu$ black holes exist as long as $2\alpha_1 p^2 \geq Q_2^2$ is sufficiently large, in particular, as long as $-|\mu|\geq (Q/2)^2[1-\log (Q/2)^2]$ holds.

As more $\alpha_n$ are turned on, the number of horizons can increase and the analysis becomes more involved ---see \eg Fig. \ref{refiss} for a couple of examples with three horizons.

As $r\rightarrow 0$, the spacetime described by \req{fgen} can look very different, depending on the value of the combination $m_{\rm max}+2-n_{\rm max}$, where we define $n_{\rm max}$ and $m_{\rm max}$ as the largest values of $n$ and $m$ corresponding to non-vanishing $\alpha_n$'s and $\beta_m$'s. In particular, the metric function goes as $f \overset{r\rightarrow 0}{\sim } r^{2(m_{\rm max}+ 2 -n_{\rm max})}$ ---where, for convenience,
if all the $\beta_m$ are zero, we define $m_{\rm max}\equiv -1$ and $\beta_{-1}\equiv -1$. 
The case $n_{\rm max}=1$ is slightly different and reads instead $f \overset{r\rightarrow 0}{\sim }r^{2(m_{\rm max}+1)} \log r$. We study the different situations in the following subsections.



\subsection{Black holes with curvature singularities}
\begin{figure}[t] \centering \vspace{-0.6cm}
	\includegraphics[scale=0.617]{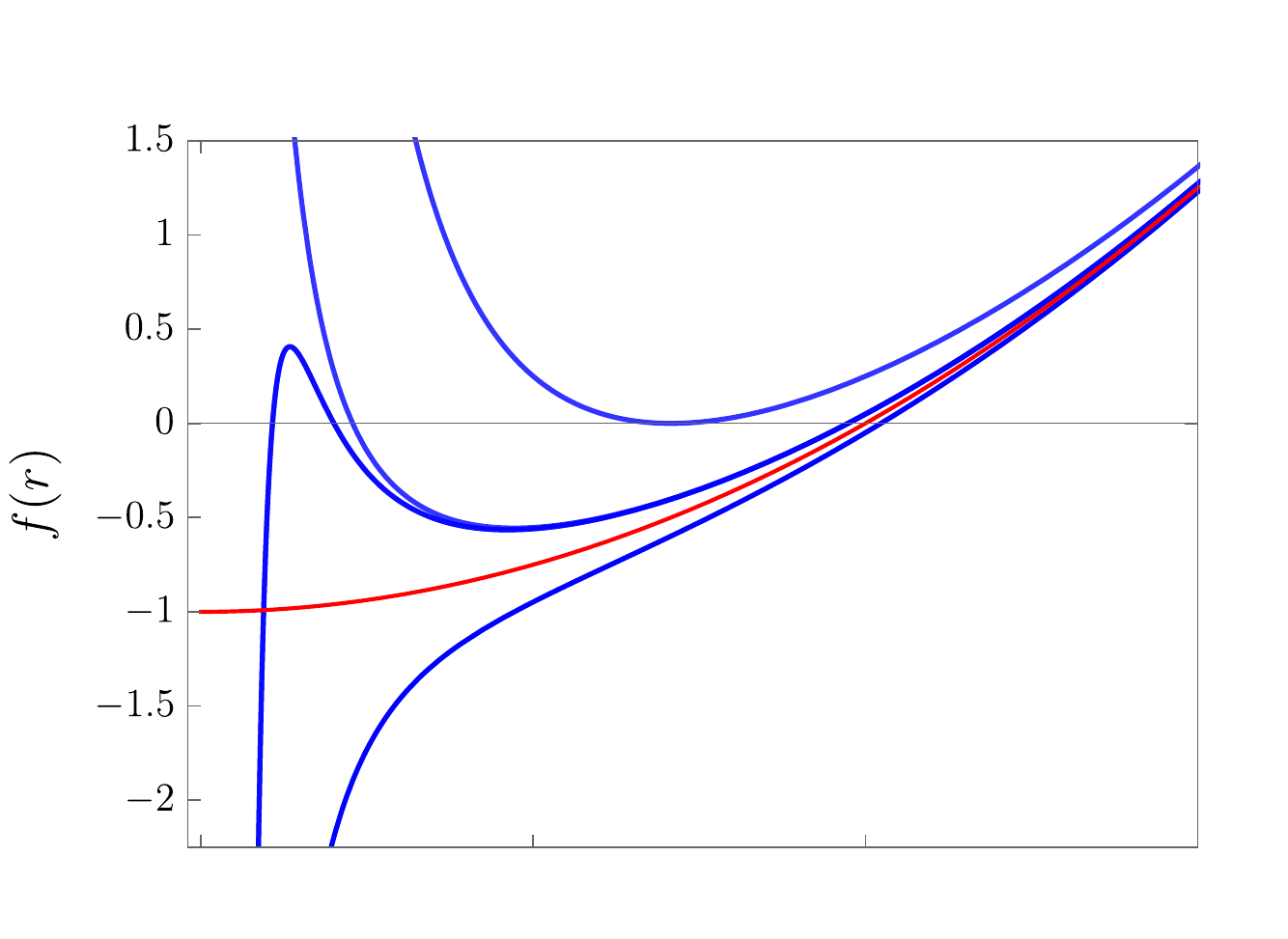}\vspace{-1.2cm}
	\includegraphics[scale=0.632]{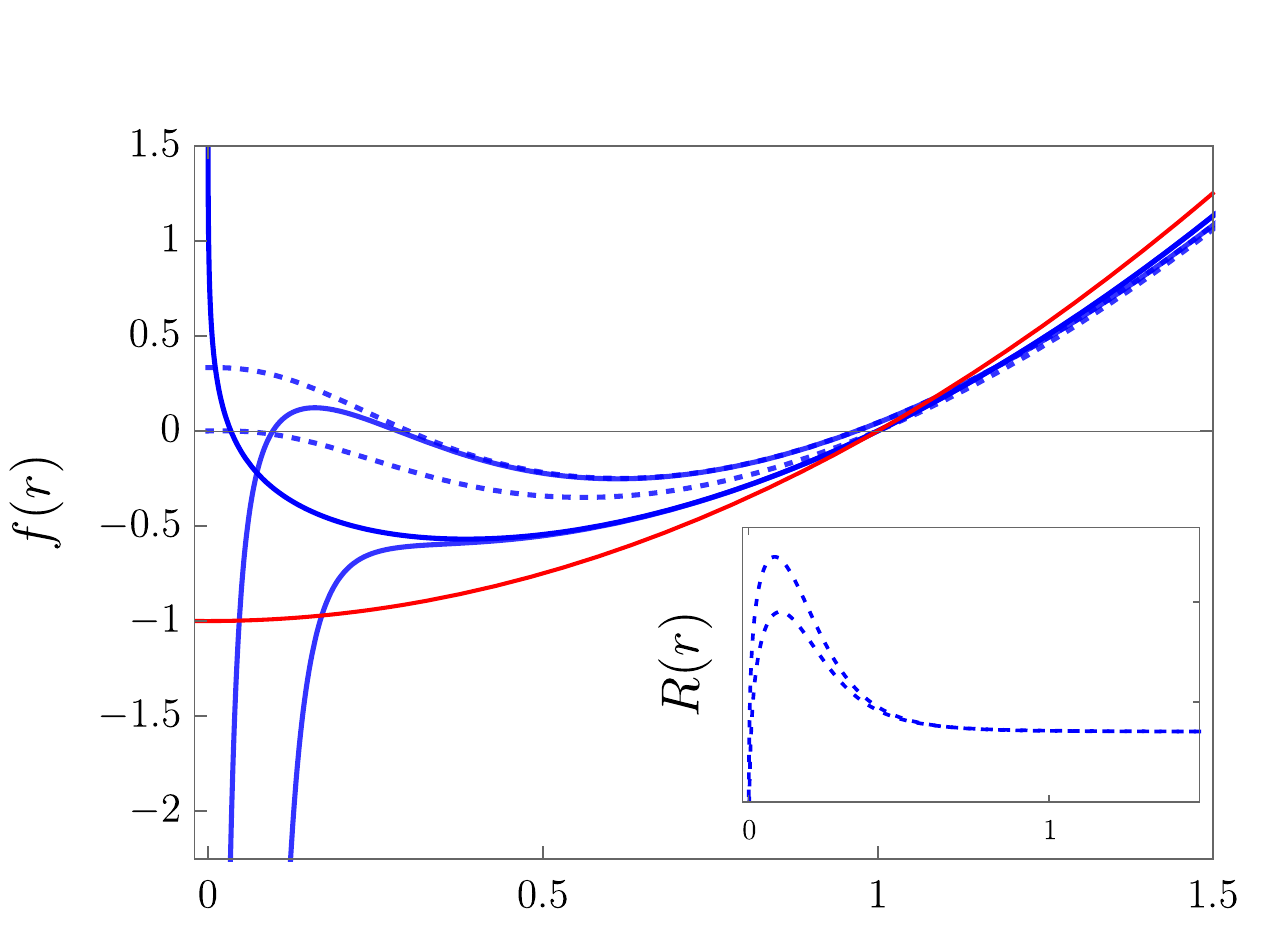}
	\caption{(Top) We plot $f(r)$ for four black hole solutions with curvature singularities at the origin possessing one, three, two and one (degenerate) horizons respectively (blue curves from bottom to top). The curves are obtained for $L=1$, $\mu=1$, $p= 2/3$ in the cases: $\alpha_2=-1/2$; $\alpha_2=1/2$, $\alpha_3=-1/50$; $\alpha_2=1/2$; and $\alpha_2=81/32$ respectively (unspecified coupling values equal zero).  
	(Bottom) We plot $f(r)$ for three black hole solutions possessing two, three and one horizon (thick blue curves moving from upper-left corner towards lower-right corner). The curves are obtained for the same values of $L,\mu,p$ in the cases: $\alpha_1=2/3$; $\alpha_1=2/3$, $\alpha_2=1/2$, $\alpha_3=-1/50$, $\beta_0=1/2$; and $\alpha_1=2/3$, $\alpha_3=-1/50$.  The dotted lines correspond to solutions for which the Ricci scalar diverges logarithmically at the origin, as shown in the inset plot. These correspond to $\alpha_1=2/3$, $\beta_0=1/2$; and $\alpha_1=2/3$, $\alpha_2=1/2$, $\beta_0=1/2$.  The red line corresponds to the usual BTZ black hole in both plots.}
	\label{refiss}
\end{figure}

An important set of solutions corresponds to black holes possessing a curvature singularity at $r=0$, hidden behind one or several horizons. This situation occurs whenever $f(r)$ contains at least a real zero, and either $n_{\rm max} > m_{\rm max} +2$ or $n_{\rm max}=1$, $m_{\rm max}=-1$ hold. We plot examples of configurations of these kinds in Fig. \ref{refiss}. Curvature invariants diverge at the origin in these cases. For instance, the Ricci scalar behaves as
\begin{equation}
R \overset{r\rightarrow 0}{= } - \frac{c_{n_{\rm max},m_{\rm max}} \alpha_{n_{\rm max}} L^{2(n_{\rm max}-m_{\rm max}-2)}}{\beta_{m_{\rm max}} r^{2(n_{\rm max} - m_{\rm max} -1)}} \, ,
\end{equation}
where $c_{n_{\rm max},m_{\rm max}}\equiv (2n_{\rm max}- 2m_{\rm max}- 5)( n_{\rm max}- m_{\rm max}-2)p^{(2n_{\rm max}-2m_{\rm max}-2)}/[(1+2m_{\rm max}) (n_{\rm max}-1)]$  is a positive constant for all $n_{\rm max}$ and $m_{\rm max}$.  

Slightly special are the cases corresponding to $n_{\rm max}=1$, $m_{\rm max}=0$ and $n_{\rm max}=2$, $m_{\rm max}=0$ (with a non-vanishing $\alpha_1$). For those, the Ricci scalar diverges logarithmically as $r\rightarrow 0$ even though $f(r)$ tends there, respectively, to zero and to a constant. The  dotted lines in Fig. \ref{refiss} correspond to these two cases.

\subsection{Black holes with BTZ-like and conical singularities}
The usual static BTZ black hole is locally equivalent to pure AdS$_3$ \cite{Banados:1992wn,Banados:1992gq}. All curvature invariants are constant and the spacetime is therefore very different at the origin from the one of the black holes considered in the previous subsection. For the BTZ, the spacetime contains a ``sort of''  conical singularity for general values of $\mu$ different from $-1$ (which precisely corresponds to pure AdS$_3$), hidden behind a horizon whenever $\mu > 0$ and naked whenever $\mu < 0$ (the $\mu=0$ case describes the so-called ``black hole vacuum'' \cite{Banados:1992gq}).  Indeed, when the mass parameter is negative and different from minus one, the $(r,\varphi)$ components of the metric have the same signature, $\diff r^2/|\mu| +r^2 \diff \varphi^2 $, which corresponds to a standard conical singularity with deficit angle $2\pi (1- \sqrt{|\mu|})$. On the other hand, when the mass parameter is positive, we have instead $-\diff r^2/|\mu| +r^2 \diff \varphi^2 $, which has a singularity in the causal structure at $r=0$ which resembles the one of a Taub-NUT space \cite{Banados:1992gq} ---in particular, the spacetime is no longer Hausdorff.

Both kinds of singularities ---conical and BTZ-like--- appear for some of the new black holes considered here. The situation described takes place for $n_{\rm max} = m_{\rm max} +2$ (with $n_{\rm max}\geq 3$ if $\alpha_1\neq 0$). In that case, the metric function and the curvature invariants tend to constant values at $r=0$. For instance, if the only active couplings are $\alpha_j$ and $\beta_{j-2}$, the metric function tends to the constant value
\begin{equation}
f\overset{r\rightarrow 0}{=}\frac{\alpha_j p^2 }{2 (j-1)(2j-3) \beta_{j-2}} \, ,
\end{equation}
whereas the Ricci scalar vanishes as  $\sim r^2$. Then, whenever the quotient $\alpha_j/\beta_{j-2}$ is positive, we have a conical singularity, and whenever it is negative, we have a BTZ-like one. The analysis is similar as more couplings are turned on. We plot examples of both kinds of solutions in the upper plot of Fig. \ref{refiss2}.


\subsection{Singularity-free black holes}
Whenever the metric function tends to $1$ at $r=0$,
\begin{equation}\label{reg11}
f(r) \overset{r\rightarrow 0}{ =}1\, ,
\end{equation}
 the angular defect present at $r=0$ disappears and the metric becomes regular there ---as mentioned earlier, this is precisely what happens with the BTZ metric for the special value $\mu=-1$, for which it reduces to pure AdS$_3$.

\begin{figure}[t!] \centering \vspace{-0.6cm}
\includegraphics[scale=0.594]{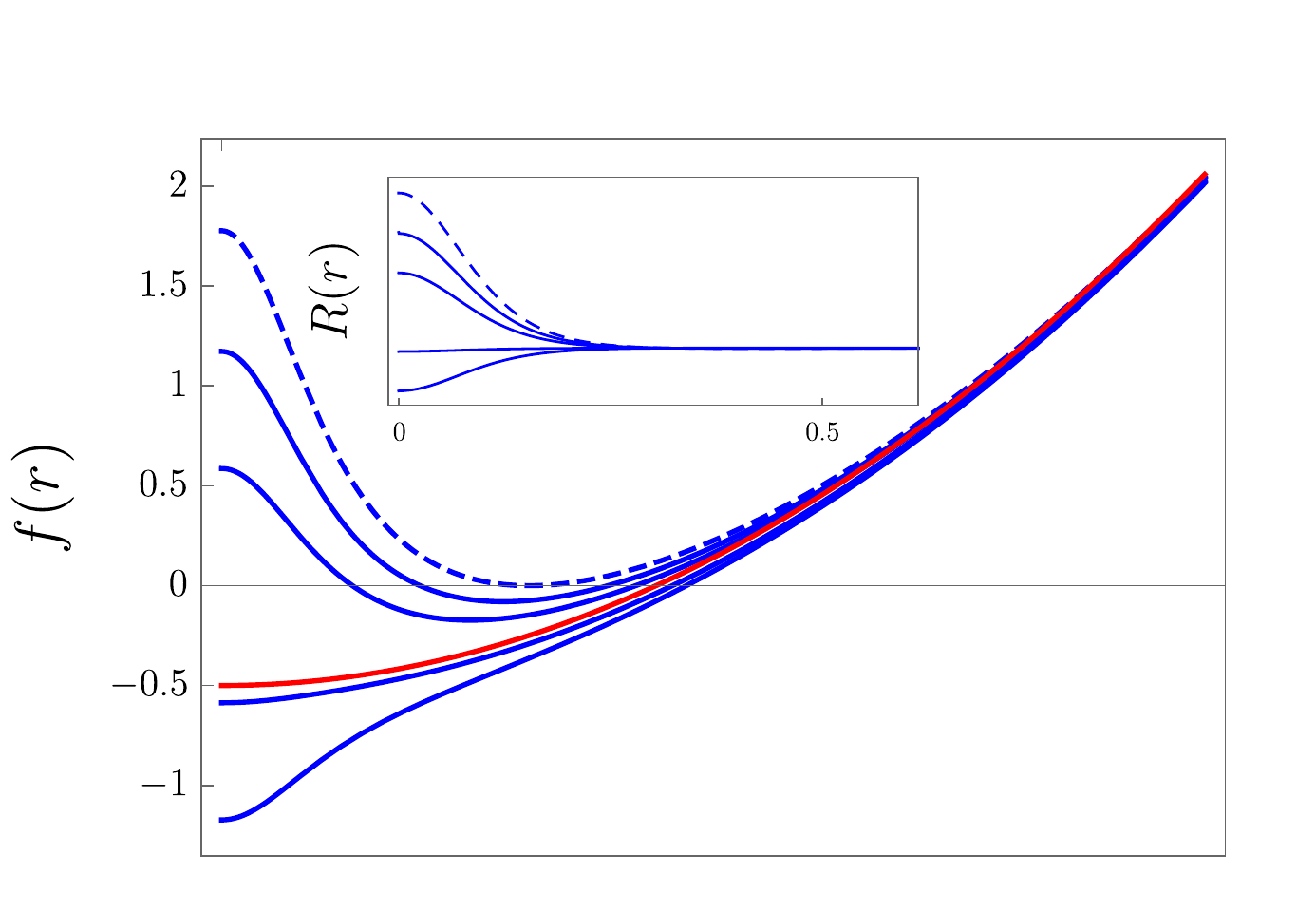}\vspace{-1cm}
	\includegraphics[scale=0.595]{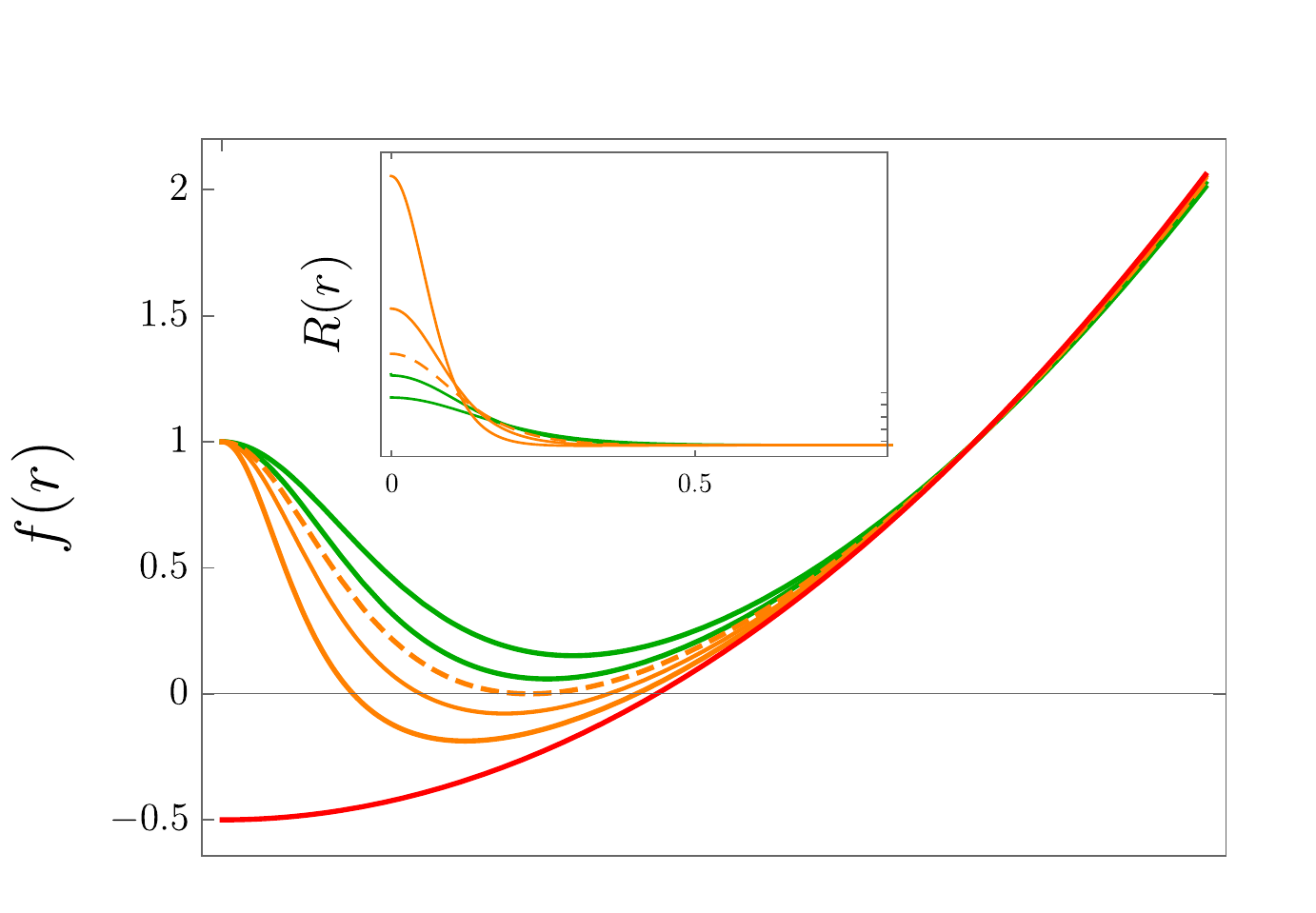}\vspace{-1cm}
	\includegraphics[scale=0.586]{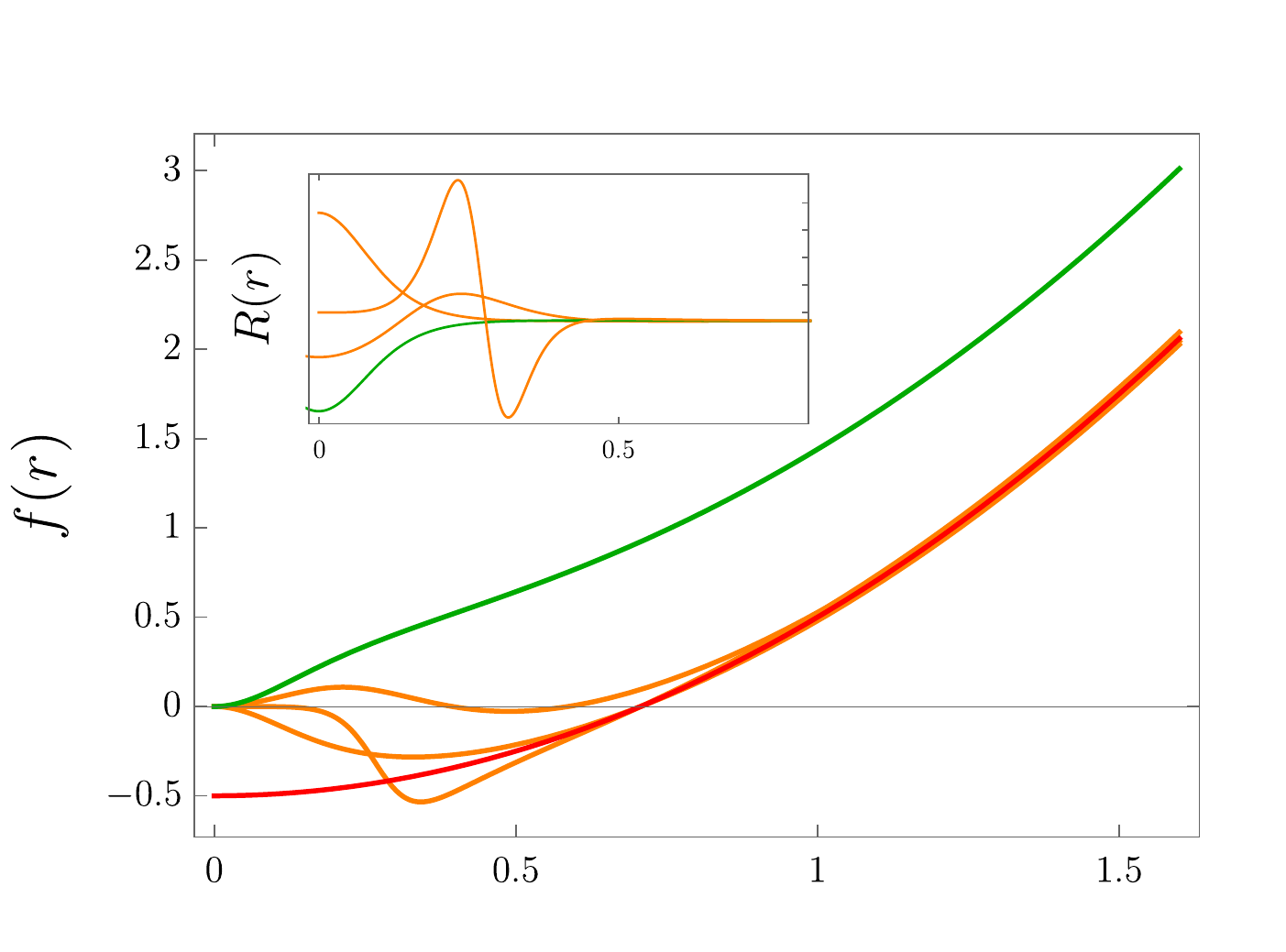}
	\caption{We plot $f(r)$ and the Ricci scalar (inset) in various cases (we set $L=1$, $\mu=1/2$, $p= 3/8$): (Top) Three black holes with a conical singularity at the origin with one (extremal), two and two horizons respectively, as well as two black holes with  BTZ-like singularities with one horizon  (blue curves from top to bottom: $\beta_0=1/4$, $\alpha_2=512/81\xi$ with $\xi=1;2/3;1/3;-1/3,-2/3$). (Middle) Two black holes with two horizons (thick orange curves), an extremal black hole (dashed) and two regular horizonless solutions (green curves): $\alpha_2=512\xi /81$ and $\beta_0=4\xi/9$ for (lower to upper): $\xi=1/3$; $\xi=2/3$; $\xi=1$; $\xi= 4/3$; $\xi=2$. (Bottom) Three regular black holes with two, one and one horizon respectively and one horizonless solution whose metric functions vanish as even powers at the origin (orange curves: $\alpha_2=54/10$, $\beta_0=8/27$, $\beta_1=1/6$; $\beta_0=8/27$; and $\beta_0=-8/27$, $\beta_1=-1/12$, $\beta_2=1/20$; and green: $\mu=-1/2$, $\beta_0=8/27$). }
	\label{refiss2}
\end{figure}


Our new solutions include non-trivial profiles for which this happens. 
When the spacetime contains at least one horizon, those describe singularity-free black holes. We plot examples of this kind  in the middle plot of Fig. \ref{refiss2}. For instance, if the only active couplings are $\alpha_j$ and $\beta_{j-2}$, the regularity condition (\ref{reg11}) becomes $\beta_{j-2}=\alpha_j p^2/[2(j-1)(2j-3)]$. The simplest example corresponds to the case $\alpha_{1,n\geq 3}=\beta_{n\geq 1}=0$, $p^2=2\beta_0/\alpha_2$. Then, the EM-QT action becomes \req{EQTG} where now
\begin{equation}\label{sing1}
\frac{\mathcal{Q}}{L^2}=\alpha_2  (\partial \phi)^4-\beta_0 [3 R^{bc}\partial_b\phi\partial_c \phi- (\partial \phi)^2R ]\, ,
\end{equation}
and the solution reads
\begin{align}\label{sing2}
f(r)=\frac{ \left[\displaystyle \frac{r^2}{L^2}-\mu +\frac{2 \beta_0^2 L^2}{ \alpha_2 r^2 }\right]}{\displaystyle \left[ 1+\frac{2 \beta_0^2 L^2}{ \alpha_2 r^2 }\right]}\, , \quad \phi = \varphi \sqrt{\frac{2\beta_0}{\alpha_2}} \, .
\end{align}
For the above metric, the Ricci scalar tends to the constant value $R=3(1+\mu)\alpha_2/( L^2 \beta_0^2)$ at the origin.
As explained at the beginning of this section ---see \req{rpm}--- this metric can describe up to two horizons  depending on the values of $2p^4 \alpha_2/\mu^2=8\beta_0^2/(\alpha_2 \mu^2)$. 

As a matter of fact, there are additional ways to achieve singularity-free black holes within the present setup which do not require imposing any constraint at all. The idea is to consider metrics for which $f(r)$ vanishes as some positive power of $r$ near the origin,
\begin{equation}\label{reg22}
f(r) \overset{r\rightarrow 0}{ =} \mathcal{O}(r^{2s})\, , \quad s\geq 1\, ,
\end{equation}
 with the curvature invariants tending to constant values there (being also finite everywhere else). This happens whenever $m_{\rm \ssc max}>n_{\rm \ssc max}-2$ if $n_{\rm \ssc max} \geq 2$; whenever some $\beta_m$ is active and all the $\alpha_n$'s are zero; and whenever $m_{\rm \ssc max} \geq 1$ if $n_{\rm \ssc max}=1$. For those, the point $r=0$ becomes a sort of new asymptotic region. The Ricci scalar behaves there as $R \sim \mathcal{O} (r^{2(m_{\rm \ssc max}-n_{\rm \ssc max}+1)})$ for $n_{\rm \ssc max} \geq 1$ and as $R \sim \mathcal{O} (r^{2m})$ if all the $\alpha_n$'s are turned off. We present examples of this kind in the bottom plot of Fig. \ref{refiss2}. There are of course infinitely many possibilities, but let us mention explicitly the simplest one. This corresponds to setting $\alpha_{n\geq 1}=0$ and $\beta_{n \geq 1}=0$. Then, the EM-QT  Lagrangian is given by \req{EQTG} where now
\begin{align}\label{sing3}
\mathcal{Q}=-\beta_0L^2 [3 R^{bc}\partial_b\phi\partial_c \phi- (\partial \phi)^2R ]\, ,
\end{align}
and the solution reads
\begin{align}\label{frl}
f(r)=\frac{ \left[\displaystyle \frac{r^2}{L^2}-\mu \right]}{\displaystyle \left[ 1+\frac{ \beta_0L^2 p^2 }{  r^2 }\right]}\, , \quad \phi= p \varphi\, , 
\end{align}
where all the constants: $\mu$, $p$, $\beta_0$ and $L^2$ are free parameters (the only conditions being $\mu>0$, $\beta_0\geq 0$, which ensure that a horizon exists and that there are no poles in the denominator). As expected, curvature invariants are finite everywhere for this solution ---see inset figure in bottom plot of Fig. \ref{refiss2}.

 In contrast with most previous attempts at achieving regular black holes, these solutions do not require any sort of: i) complicated functional dependences of the action fields; ii) fine tuning of action parameters or constraint between those and the physical charges; iii) addition of specially selected matter. For the Lagrangian density in \req{EQTG} with $\mathcal{Q}$ given by \req{sing3}, black holes simply turn out to be singularity-free ---and analogously in the rest of the cases described. 


\subsection{Regular horizonless solutions}
Solutions behaving as \req{reg11} or as \req{reg22} with finite-everywhere curvature invariants do not necessarily include horizons. When they do not, we are left instead with globally regular and horizonless spacetimes. For example, the solutions in  \req{sing2} and in \req{frl} with $\mu \leq 0$ describe configuration of this kind. Examples of regular horizonless solutions are shown in green in the bottom and middle plots of Fig. \ref{refiss2}.  



\section{Adding rotation}\label{rotat}
The static solutions described by \eqref{eq:singlefmetric} can be easily generalized into rotating ones by performing a boost in the $t$ and $\varphi$ coordinates,
\begin{equation}
t\rightarrow \gamma t-L\omega \varphi\, , \quad \varphi\rightarrow \gamma\varphi -\omega t/L\, .
\end{equation}
The ``trick'' is that this change of variables is only defined locally, so that the global structure of the resulting spacetimes is different from their static counterparts ---see \eg  \cite{Dias:2002ps,CamposDias:2003tv}. Assuming that $\gamma^2-\omega^2=1$ ---so that for $\omega=0$ the metric reduces to the static one---  we get
\begin{align}\label{eq:rotatingmetric}
\diff s^2&=-\frac{r^2f}{\rho^2}\diff t^2+\frac{\diff r^2}{f}+\rho^2\left[\diff \varphi+N^{\varphi}\diff t\right]^2\, ,\\ \phi&=p\left[ \gamma\varphi-\frac{\omega t}{L}\right]\, ,
\end{align}
where
\begin{align}
\rho^2 &\equiv r^2-\omega^2[L^2f-r^2]\, ,\\ N^{\varphi} &\equiv \frac{\gamma\omega [L^2f -r^2]}{L \rho^2}\, .
\end{align}
The rotating solution typically has the same horizons as the static one plus an additional one at $r=0$ if $f(r)$ tends to a non-vanishing constant there. 


\section{Final comments}\label{finalco}
In this letter we have put forward a new family of three-dimensional gravity theories involving a non-minimally coupled scalar field. These ``Electromagnetic Quasitopological'' theories possess static solutions ---easily generalizable to rotating ones, as shown in Section \ref{rotat}--- characterized by a single function $f(r)$ whose general form appears in \req{fgen} above. These describe different kinds of solutions depending on the values of the various couplings and parameters. In particular, we have argued that black holes with different numbers of horizons and conical, BTZ-like or curvature singularities appear in some cases. In others, singularities are absent from the geometry and the solutions describe singularity-free black holes or globally regular solutions. This is achieved in two ways, characterized by the behavior of $f(r)$ near the origin ---see \req{reg11} and \req{reg22} respectively and orange curves in the middle and bottom plots of Fig. \ref{refiss2}. Interestingly, for the second type of solutions, this behavior is encountered without imposing any kind of constraint between the action parameters and/or physical charges.

There are many venues we consider worth exploring regarding the solutions presented here. On the one hand, it would be interesting to study their geodesic structure and Penrose diagrams. A related issue is a more detailed characterization of the structure and nature of the horizons and singularities described by the solutions ---the tools developed in \cite{OliveiraNeto:1996qm,Tavlayan:2020chf} could be useful for this. It would also be interesting to explore the thermodynamic properties of the solutions, as well as possible holographic applications. 

On a different front, it would be interesting to determine whether additional EM-QT theories exist in three dimensions and to study the case of EM-GQT theories, for which the requirement that the equation of $f(r)$ is algebraic is relaxed.

\begin{acknowledgments}   We thank  \'Angel Murcia, Julio Oliva and Bayram Tekin for useful discussions. The work of  PB was supported by the Simons Foundation through the ``It From Qubit'' Simons collaboration.  The work of PAC is supported by a postdoctoral fellowship from the Research Foundation - Flanders (FWO grant 12ZH121N). The work of JM is funded by the Agencia Nacional de Investigaci\'on y Desarrollo (ANID) Scholarship No. 21190234 and
by Pontificia Universidad Cat\'olica de Valpara\'iso. JM is also grateful to the QMAP faculty for their hospitality. The work of GvdV is supported by CONICET and UNCuyo, Inst. Balseiro.
\end{acknowledgments}

\onecolumngrid  \vspace{1cm} 
\appendix 

\section{Equations of motion} \labell{eqsofmo}
In this appendix we present the full equations of motion of the EM-QT theory introduced in the main text. We also present an on-shell approach in which we insert the ansatz
\begin{equation}\label{Nf}
\diff s^2=-N^2(r)f(r)\diff t^2+\frac{\diff r^2}{f(r)}+r^2\diff \varphi^2\, , \quad \phi= \phi(\varphi)\, ,
\end{equation}
directly in the action and vary with respect to $N$, $f$ and $\phi$. The equations obtained for these functions from this approach are equivalent to the full non-linear equations evaluated in the same ansatz. The equations of $f$ and $\phi$ are then simply solved by $N=1$ and $\phi = p \varphi$, as anticipated in the main text. The remaining one, obtained from the variation with respect to $N$ in the action approach, can be integrated once and yields the algebraic equation of $f(r)$ which appears in the main text.


Let us start with the  action of our family of EM-QT theories, 
\begin{equation}
I_{\ssc \rm EMQT}=\frac{1}{16\pi G}\int\diff^3x\sqrt{|g|}\left[R+\frac{2}{L^2}-\mathcal{Q} \right]\, ,
\end{equation}
where
\begin{equation}
 \mathcal{Q}  =\sum_{n=1} \alpha_n L^{2(n-1)} (\partial\phi)^{2n}-\sum_{m=0} \beta_m L^{2(m+1)}(\partial\phi)^{2m}\left[ (3+2m) R^{cd} \partial_c \phi \partial_d \phi- (\partial\phi)^2 R \right] \, .
\end{equation}
For the variation with respect to the inverse metric tensor $g^{ab}$ we find
\begin{align}
&\frac{1}{\sqrt{|g|}}\frac{\delta I_{\ssc \rm EMQT}}{\delta g^{ab}}=G_{ab}-\frac{1}{L^2}g_{ab}-\sum_{n}\alpha_n (\partial\phi)^{2(n-1)}L^{2(n-1)}\left[n\partial_a\phi\partial_b\phi-\frac{1}{2}g_{ab}(\partial\phi)^2\right]\\\notag
&\quad +\sum_{m}\beta_{m}L^{2(m+1)}(\partial\phi)^{2(m-1)}\left[2(3+2m)R_{c(a}\partial_{b)}\phi\partial^{c}\phi (\partial\phi)^2+m(3+2m)\partial_a\phi\partial_b\phi R^{cd}\partial_{c}\phi\partial_{d}\phi \right. \\
&\quad \left.  -R_{ab}(\partial\phi)^4-(m+1)\partial_{a}\phi\partial_{b}\phi(\partial\phi)^2 R\right]-\frac{1}{2}g_{ab}R^{cd}Q_{cd}-\nabla^{c}\nabla_{(a}Q_{b)c}+\frac{1}{2}\nabla^2Q_{ab}+\frac{1}{2}g_{ab}\nabla_{cd}Q^{cd}\, ,\notag
\end{align}
where
\begin{equation}
G_{ab}\equiv R_{ab}-\frac{1}{2}g_{ab}R,\quad Q_{ab}\equiv \sum_{m}\beta_{m}L^{2(m+1)}(\partial\phi)^{2m}\Big[(3+2m)\partial_{a}\phi\partial_{b}\phi-g_{ab}(\partial\phi)^2\Big]\, .
\end{equation}
The variation with respect to the scalar reads in turn
\begin{align}
&\frac{1}{\sqrt{|g|}}\frac{\delta I_{\ssc \rm EMQT}}{\delta \phi}=2\nabla_a \left[ \sum_{n=1}n\alpha_n L^{2n-1}(\partial\phi)^{2(n-1)}\partial^a\phi \right. \\ \notag
& \left. -\sum_{m=0}\beta_mL^{2(m+1)}(\partial\phi)^{2(m-1)}\left[m(3+2m)\partial^a\phi R^{bc}\partial_b\phi\partial_c\phi+(3+2m)(\partial\phi)^2R^{ab}\partial_b\phi-(m+1)R(\partial\phi)^2\partial^a\phi\right]\right]\, .
\end{align}
Now, it is possible to verify that a solution of the form \req{Nf} with $\phi = p \varphi$, $N(r)=1$ and $f(r)$ given by \req{fgen} solves all the components of the above equations.

An alternative route which turns out to be equivalent involves considering the on-shell effective Lagrangian  $L_{f,N,\phi}=\sqrt{|g|}\mathcal{L} |_{\text{(\ref{Nf})}}$ and taking variations with respect to the undetermined functions, namely 
\begin{equation}
\mathcal{E}_N\equiv\frac{\delta L_{N,f,\phi}}{\delta N}\, ,\quad \mathcal{E}_f\equiv\frac{\delta L_{N,f,\phi}}{\delta f}\, , \quad \mathcal{E}_\phi\equiv\frac{\delta L_{N,f,\phi}}{\delta \phi}\, .
\end{equation}
For each of them we find
\begin{align}
\mathcal{E}_N =&\frac{2r}{L^2}-\sum_{n=1}\frac{\alpha_n (\dot \phi)^{2n}L^{2(n-1)}}{r^{2(n-1)}}-f'+\sum_{m=0}\frac{\beta_m(\dot \phi)^{2(m+1)}(2m+1)L^{2(m+1)}}{r^{2m+3}}\left[2(m+1)f-rf'\right]=0\, ,\\
\mathcal{E}_f= &N'\cdot \left[1+\sum_{m=0}\frac{\beta_m(\dot \phi)^{2(m+1)}(2m+1)L^{2(m+1)}}{r^{2(m+1)}}\right]=0\, ,\\
\mathcal{E}_\phi =&\ddot \phi \cdot \Bigg[ N\sum_{n=1}\frac{2n (2 n-1)(\dot \phi)^{2(n-1)}\alpha_n  L^{2 (n-1)} }{r^{2(n-1)}} + \sum_{m=0} \frac{2(m+1) (2 m+1)\beta_mL^{2 (m+1)}}{r^{2 (m+1)}}( \dot \phi)^{2m}   \\ 
&\left(N' \left[-3 r f'+(2 m
   +1)f\right]+N \left[(2 m+1) f'-r f''\right]-2 r f N''\right) \Bigg]=0\, ,\notag
\end{align}
where we used the notation $g'\equiv \diff g/ \diff r$ and $\dot \phi \equiv \diff \phi/ \diff \varphi $. Now, it is evident that a linear dependence on the angle for $\phi$ and a constant one for $N(r)$ automatically solve the last two equations. On the other hand, the first equation can be integrated once to obtain the algebraic equation for $f(r)$ appearing in \req{fgen}. 
Note that there is no dependence at all on $N(r)$ in such equation, which means that the solutions with $N(r)=$ constant, implied by the equation of $f(r)$, are the only possible ones. This is a general property of Quasitopological gravities (Electromagnetic or not).


\bibliographystyle{apsrev4-1} 
\vspace{1cm}
\bibliography{GravitiesINSPIRE2}

\end{document}
%